\documentclass{article}
\usepackage{spconf,amsmath,graphicx}

\usepackage{enumitem}
\setlist{nosep, leftmargin=14pt}
\usepackage{booktabs}
\usepackage{mwe} % to get dummy images

\title{Hybrid Diffusion Model for Breast Ultrasound Image Augmentation}
\name{
Farhan Fuad Abir$^{1, 2}$ \qquad
Sanjeda Sara Jennifer$^{3}$ \qquad
Niloofar Yousefi$^{4}$ \qquad
Laura J. Brattain$^{2}$
}

\address{
$^{1}$Electrical and Computer Engineering, University of Central Florida \\
$^{2}$Department of Medicine, University of Central Florida College of Medicine \\
$^{3}$Department of Computer Science, University of Central Florida \\
$^{4}$Industrial Engineering and Management Systems, University of Central Florida
}

\begin{document}
%\ninept
%
\maketitle
\begin{abstract}
We propose a hybrid diffusion-based augmentation framework to overcome the critical challenge of ultrasound data augmentation in breast ultrasound (BUS) datasets. Unlike conventional diffusion-based augmentations, our approach improves visual fidelity and preserves ultrasound texture by combining text-to-image generation with image-to-image (img2img) refinement, as well as fine-tuning with low-rank adaptation (LoRA) and textual inversion (TI). Our method generated realistic, class-consistent images on an open-source Kaggle breast ultrasound image dataset (BUSI). Compared to the Stable Diffusion v1.5 baseline, incorporating TI and img2img refinement reduced the Fréchet Inception Distance (FID) from 45.97 to 33.29, demonstrating a substantial gain in fidelity while maintaining comparable downstream classification performance. Overall, the proposed framework effectively mitigates the low-fidelity limitations of synthetic ultrasound images and enhances the quality of augmentation for robust diagnostic modeling.
\end{abstract}
\section{Introduction}
Ultrasound is a key medical imaging modality for breast cancer diagnosis due to its affordability, non-invasive nature, and real-time imaging capabilities \cite{khalid2023breast}. However, deep learning–based ultrasound interpretation faces persistent challenges due to limited and imbalanced datasets that constrain model generalizability across lesion types. Traditional augmentation techniques, such as flipping, rotation, and intensity variations, offer only limited diversity \cite{Goceri2023MedicalAugmentation}. Early generative approaches using Generative Adversarial Networks (GANs) often produce artifacts and fail to replicate the speckle noise and subtle tissue textures critical for diagnosis \cite{jimenez2024gan}.

Diffusion models progressively denoise latent representations, yielding structural detail essential for ultrasound synthesis. These models have been widely used for denoising, despeckling \cite{zhang2023adding, asgariandehkordi2024denoising} and text-conditioned generation of echocardiograms or breast ultrasound (BUS) images \cite{freiche2025ultrasound, oh2024breast}. These methods showed promising results of using diffusion-based image generation for BUS. However, existing diffusion-based BUS generation tends to produce overly smooth images that lack the characteristic speckle noise and structural complexity of real BUS images.

To address the research gap, we propose a hybrid diffusion augmentation approach to improve the fidelity of ultrasound synthesis and the severe class imbalance in breast ultrasound datasets. Our framework integrates semantic conditioning with text-to-image (text2img) generation and structural refinement with image-to-image (img2img) diffusion \cite{rombach2022high}. First, we implemented Stable Diffusion v1.5 (SD1.5) \cite{rombach2022high}, fine-tuned with low-rank adaptation (LoRA) \cite{hu2022lora} and enhanced with textual inversion (TI) \cite{gal2022image} to embed domain-specific ultrasound concepts into the model’s latent space. Then, we use img2img for retaining fine-grained BUS structures. This hybrid strategy retains ultrasound texture, tissue heterogeneity, and lesion boundaries more faithfully than other diffusion-based approaches. We used the proposed framework to generate synthetic samples for underrepresented classes, addressing data imbalance in the open-source Kaggle breast ultrasound image (BUSI) dataset \cite{al2020dataset}. By producing class-consistent and anatomically coherent ultrasound images, the method improves downstream classification performance.

\section{Methods}

\begin{figure*}[thb]
  \centering
  \includegraphics[width=0.8\textwidth]{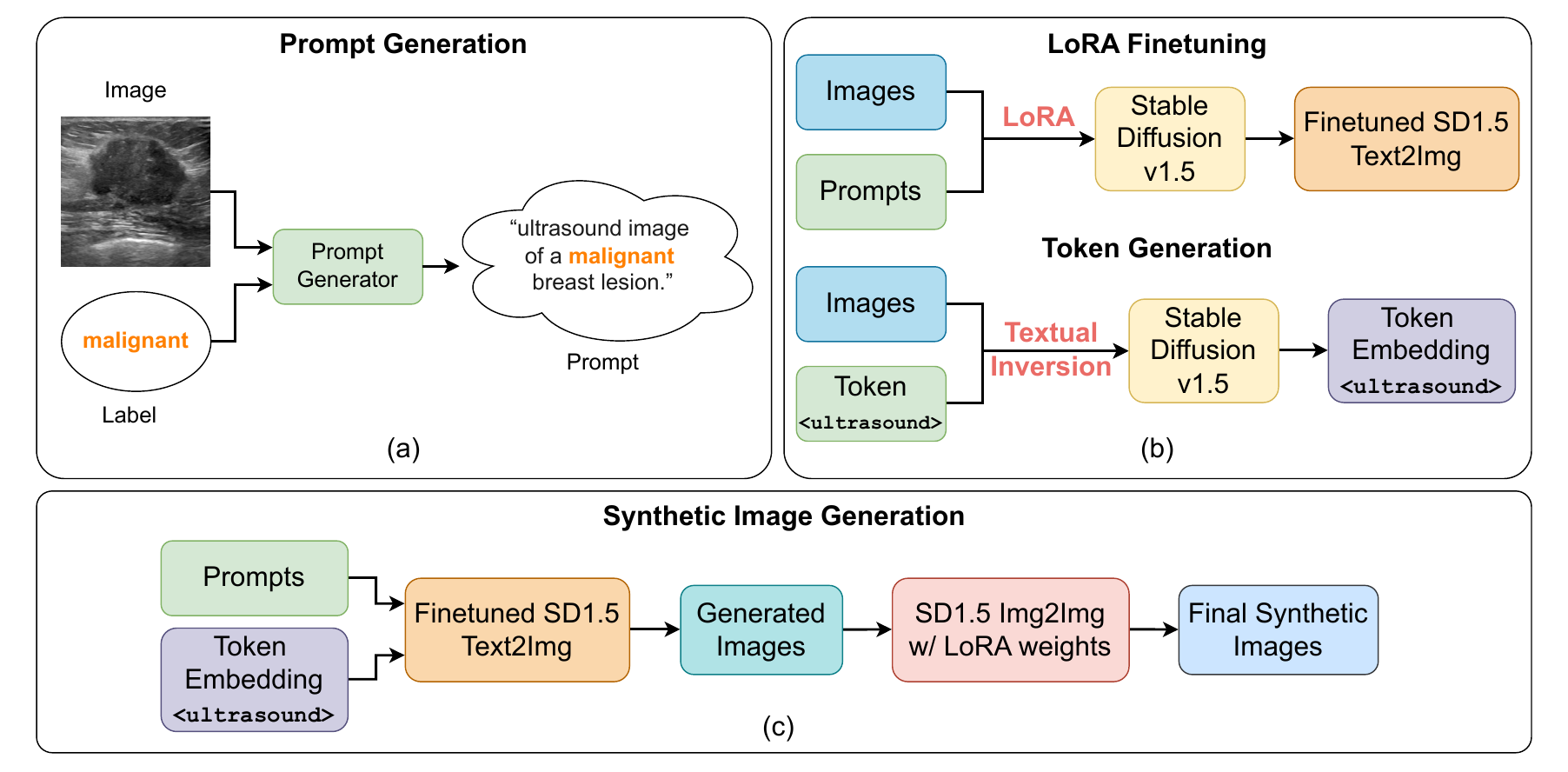}
  \caption{Overview of the proposed hybrid diffusion-based image generation framework for breast ultrasound. The method consists of three main stages. (a) The preprocessing stage converts the labels into descriptive prompts. (b) LoRA finetuning and Token Generation adapts the Stable Diffusion v1.5 using LoRA for image–prompt alignment and Textual Inversion for learning domain-specific \texttt{<ultrasound>} token. (c) In the final workflow, the prompts and learned token \texttt{<ultrasound>} are passed through the finetuned text-to-image (text2img) model to generate synthetic images, which are further refined using an image-to-image (img2img) stage with LoRA weights, yielding the final synthetic ultrasound images.}
  \label{fig:methodology}
\end{figure*}

\subsection{Dataset Description}
We used the Kaggle open-source Breast Ultrasound Images (BUSI) dataset consisting of 780 images categorized as benign (437), malignant (210), and normal (133) \cite{al2020dataset}. All images were resized and normalized. The train-validation split (80-20) had 623 training images (349 benign, 168 malignant, 106 normal) and 157 validation images (88 benign, 42 malignant, 27 normal), with benign being the majority class.

\subsection{Hybrid Diffusion Model}

Fig.\ref{fig:methodology} illustrates the three-stage model workflow: prompt generation, model fine-tuning, and synthetic image generation. 

\textbf{Prompt Generation.} We derived class-specific prompts from BUSI labels (normal, benign, malignant) to guide the generation process. These prompts were mapped to radiology-style descriptions such as “ultrasound image of a benign (or malignant) breast lesion.” For the normal class, the prompt was: “ultrasound image of normal breast tissue.” 

\textbf{Model Fine-Tuning.} Firstly, we used low-rank adaptation (LoRA) to fine-tune the Stable Diffusion model with the images and associated text prompts. This facilitated domain-specific feature learning while avoiding full-scale model pretraining. Then we introduced a custom token \texttt{<ultrasound>} using TI to enhance the model’s understanding of ultrasound-specific prompts. TI learned a new embedding that captured characteristic texture and speckle patterns from representative BUSI images. This embedding was incorporated into all prompts to strengthen domain alignment and ensure consistent synthesis across BUS classes.

\textbf{Synthetic Image Generation.} At first, the TI embedding was imported alongside the LoRA adapters. Then we generated the synthetic BUS images from the finetuned text2img SD1.5 model. Additionally we applied Stable Diffusion’s img2img model to enhance the fidelity of the generated outputs. This refinement process was performed with an empirically chosen low denoising strength of 0.3. Afterward, we added synthetic images to each class of the training set, yielding 350 samples per class to match the majority class, while the validation set remained unchanged.

\subsection{Evaluation Metrics} 

We used accuracy, F1-score, Area Under the Receiver Operating Characteristic Curve (AUC-ROC), positive predictive value (PPV), and recall to evaluate classification performance. Moreover, we used the Fréchet Inception Distance (FID), which quantifies the similarity between real and synthetic image distributions. FID scores were computed using Inception v3 features on 780 real and synthetic images.

\begin{table*}[!tbp]
\centering
\caption{Impact of each component in the hybrid diffusion framework on classification and image quality metrics.}
\label{tab:results}
\begin{tabular}{lccccc}
\toprule
\textbf{Components} & \textbf{Accuracy} $\uparrow$ & \textbf{F1-Score} $\uparrow$ & \textbf{AUC-ROC} $\uparrow$ & \textbf{PPV} $\uparrow$ & \textbf{FID} $\downarrow$ \\
\midrule
Baseline (Original Images)                     & 0.904 & 0.887 & 0.979 & 0.890 & -       \\
Original + SD1.5                  & 0.917 & 0.905 & \textbf{0.986} & 0.901 & 45.97   \\
Original + SD1.5 + img2img        & 0.898 & 0.879 & 0.978 & 0.878 & 38.34   \\
Original + SD1.5 + TI             & \textbf{0.924} & \textbf{0.912} & 0.980 & \textbf{0.906} & 45.66   \\
Original + SD1.5 + TI + img2img   & 0.905 & 0.884 & 0.975 & 0.89 & \textbf{37.18}   \\
\bottomrule
\end{tabular}
\end{table*}

\begin{figure*}[!ht]
  \centering
  \includegraphics[width=0.8\textwidth]{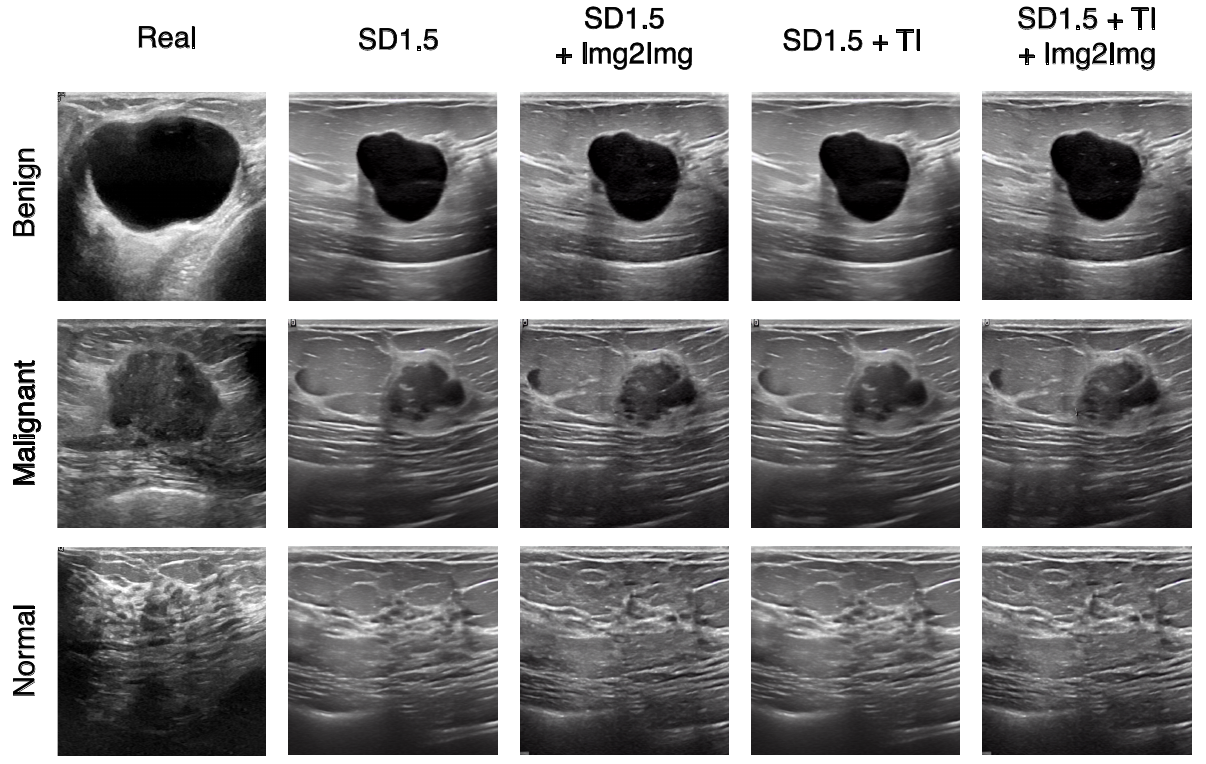}
  \caption{Comparison of real ultrasound images and synthetic variants generated by 4 different approaches based on SD1.5. Rows correspond to breast lesion categories: benign (top), malignant (middle), and normal (bottom). Columns show (from left to right): real images, baseline SD1.5 generations, SD1.5 with img2img refinement, SD1.5 with TI, and SD1.5 combined with TI and img2img. The img2img refinement increases the fidelity by improving ultrasound textures.}
  \label{fig:generated_images}
\end{figure*}

\section{Experiments and Results}

\subsection{Experimental Setup} 
We conducted five experiments using original ultrasound images as the baseline. ResNet18 was used to classify the three classes. (1) Baseline (Original): trained only on original ultrasound images; (2) Original + SD1.5: augmented with synthetic samples generated by SD1.5; (3) Original + SD1.5 + img2img: adding img2img translation for style adaptation; (4) Original + SD1.5 + TI: leveraging textual inversion for domain-specific embedding alignment; and (5) Original + SD1.5 + TI + img2img: combining all components.

ResNet18 was trained with the Adam optimizer (learning rate = 0.0001, batch size = 16, 30 epochs) using cross-entropy loss. We fine-tuned SD1.5 using the standard denoising objective whereas for TI we updated the learned token embedding. Standard augmentations (horizontal flip, normalization) were applied to improve generalization. All experiments ran on a workstation with a 12th Gen Intel Core i5-12500 CPU, 128 GB RAM, and an NVIDIA RTX A4000 (16 GB) GPU on Ubuntu Linux. Model training and diffusion-based synthesis were implemented with Python, PyTorch \cite{he2016deep}, and the Hugging Face diffusers library \cite{von-platen-etal-2022-diffusers}.

\subsection{Results}

Table \ref{tab:results} summarizes the impact of LoRA fine-tuning, TI, and img2img refinement. Using SD1.5 text2img augmentation with real data improved the baseline to Acc 0.917, F1 0.905, and the highest AUC-ROC 0.986 (FID 45.97). Incorporating TI achieved the best overall classification (Acc 0.924, F1 0.912, PPV 0.906), demonstrating that the learned domain token enhances class-aware synthesis (FID 45.66). Img2img consistently lowered FID (to 38.34 without TI and 37.18 with TI) but slightly reduced accuracy and F1. AUC-ROC remained $\ge$ 0.975 across all settings, indicating stable discriminative performance. Overall, the classifier achieved comparable classification performance on the original imbalanced data and the synthetic dataset. However, our method shows significant decrease in FID score which highlights the improved visual fidelity of the generated images.

\subsection{Qualitative Analysis}

Fig.\ref{fig:generated_images} shows examples of synthetic BUS images of 3 different classes. The images demonstrate that finetuned SD1.5 captures ultrasound features, such as lesion boundaries, internal textures, and background anatomy. Malignant lesions display irregular, heterogeneous texture, while benign ones exhibit smoother contours with homogeneous texture. Normal cases show no lesions and variable breast tissues. In addition, img2img refinement helps preserve fine-grained tissue patterns relative to SD1.5, while textual inversion enhances semantic consistency. Overall, the results suggest the synthetic images retain clinically meaningful traits. This visual validation complements the quantitative improvements, highlighting the capability of the proposed method for generating high-fidelity BUS images.

\section{Conclusion}

We presented a hybrid diffusion–based augmentation framework for breast ultrasound images that integrates prompt-driven text2img synthesis with LoRA finetuning and Textual Inversion, with an img2img refinement stage. Applied to the Kaggle BUS dataset (BUSI), our method generated high-fidelity breast ultrasound images. The refinement stage significantly improved visual quality, retaining ultrasound characteristics relevant for diagnosis. Although the small dataset failed to capture the data imbalance issue, the proposed method achieved comparable classification performance and substantially improved the FID score.

Future work will focus on closing this gap via task-aware generation, such as conditioning on lesion masks or structure-preserving priors, jointly optimizing with a diagnostic encoder, and weighting synthetic samples by classifier confidence. We will also further refine our approach on larger datasets. Finally, we will assess generalizability to other medical imaging modalities, such as MRI, CT, and X-ray.

\section{Compliance with ethical standards}
\label{sec:ethics}
This study uses the publicly available Breast Ultrasound Images (BUSI) \cite{al2020dataset}, which is fully anonymized and does not contain any personally identifiable information. Therefore, no additional ethical approval was required for this work.

\section{Acknowledgements}
\label{sec:acknowledgement}
No funding was received for this study, and the authors have no relevant financial or non-financial interests to disclose.

\bibliographystyle{IEEEbib}
\bibliography{refs}

\end{document}